\documentclass[twocolumn,preprintnumbers,amsmath,amssymb,prl]{revtex4}
\usepackage{graphicx}

\newcommand{\ket}[1]{\vert#1\rangle}

\begin{document}

\title{Manipulating time-bin qubits with fiber optics components}

\author{F\'elix Bussi\`eres}
\altaffiliation[Also at ]{Laboratoire d'informatique th\'eorique et quantique,
Universit\'e de Montr\'eal, C.P. 6128, Succ. Centre-Ville, Montr\'eal (QC), H3C~3J7 Canada}
\email{felix.bussieres@polymtl.ca}
\author{Yasaman Soudagar}
\author{Guido Berlin}
\author{Suzanne Lacroix}
\author{Nicolas Godbout}
\affiliation{Centre d'optique, photonique et lasers, Laboratoire des fibres optiques, D\'epartement de g\'enie physique, \'Ecole Polytechnique de Montr\'eal, C.P. 6079, Succ. Centre-ville, Montr\'eal (QC), H3C~3A7 Canada}

\begin{abstract}
We propose two experimental schemes to implement arbitrary unitary single qubit operations on single photons encoded in time-bin qubits. Both schemes require fiber optics components that are available with current technology.
\end{abstract}

\maketitle

\begin{figure}[!h]
\includegraphics[scale=0.7]{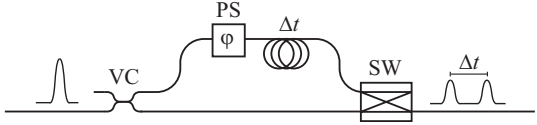}
\caption{\label{fig:time-bin} Experimental set-up to generate time-bin qubits. Lines are optical fibers, VC is a variable coupler, PS is a phase shifter and SW is an optical switch.}
\end{figure}

Photons guided in optical fibers are natural candidates to implement long distance quantum communication tasks such as quantum cryptography, quantum repeaters, testing Bell's theorem and quantum teleportation \cite{BB84,BDCZ98,Bell64,BBCJPW93}. For these purposes, polarization might seem to be an obvious way to encode qubits, but over long distances, polarization mode dispersion (PMD) and random birefringence fluctuations induce decoherence and an irreversible loss of information. To overcome this problem, time-bin coding has been proposed \cite{BGTZ99} and used in a few recent quantum communication experiments \cite{GRTZ02,TW01,MRTZG03}. To generate a time-bin qubit, a single photon pulse is split in two components using a variable coupler (VC) which is the all-fiber equivalent of a variable beam-splitter (see~FIG.~\ref{fig:time-bin}). 
Each component travels through either the ``short'' or ``long'' arm of the interferometer and then both are recombined into the same fiber using an optical switch (SW). At the output, the single photon is in a coherent superposition of being in two mutually exclusive time-bins separated by $\Delta t$: $\ket{\psi} = \alpha \ket{s} + \mathrm{e}^{i\varphi} \beta\ket{l}$, where $\ket{s}$ and $\ket{l}$ stand for ``short'' and ``long''. By changing the coupling ratio of VC and the phase delay of the phase shifter (PS), one can select any values for $\alpha$ and $\varphi$ ($\beta$ being fixed by normalization). Time-bin coding has been used to efficiently generate entanglement that is robust against PMD, random birefringence and chromatic dispersion over 50~km of standard fiber~\cite{MRTZLG04,FGRZ04}. In what follows, we will assume that we are working with wavelengths around $\lambda \approx 1.55~\mu$m so that the loss is minimal in standard single mode fiber.

The main difficulty with time-bin coding is that both single qubit unitary operations and deterministic measurements in arbitrary basis are not trivial to implement \footnote{By ``single qubit unitary operation'' we mean any 2$\times$2 unitary evolution operator applied to a given state. Measuring in an arbitrary basis is done by applying the operation that maps the basis to be measured to the eigenbasis of the detector. For time-bin qubits, measuring in the eigenbasis simply means monitoring the arrival time of the photon.}. Instead, most experiments use non-deterministic operations with post-selection~\cite{GRTZ02,MRTZG03,TW01} and this inevitably reduces the success rate and may also lead to a failure. 
For example, completely de\-ter\-mi\-nis\-tic quantum teleportation of a time-bin qubit can not be achieved wi\-thout using deterministic single qubit operations to correct the teleported state, as it has been done recently~\cite{MRTZG03}. The ubiquitous need for single qubit operations in quantum communication and the lack of any general scheme to do this are the motivations of our work. In this paper, we propose two different experimental schemes to perform arbitrary operations on time-bin qubits. Both schemes are very simple to implement using technology that is currently available.

All rotations of polarization qubits are easy to implement both in free-space and in optical fibers using waveplates or their all-fiber equivalent. The first scheme we propose takes advantage of this by converting a time-bin qubit to a polarization qubit to perform a polarization rotation and then converting back to a time-bin qubit thereafter.
\begin{figure}[!h]
\includegraphics[scale=0.7]{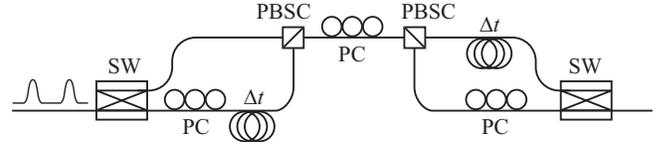}
\caption{\label{fig:polarization-gate} Experimental layout of the time-bin gate using polarization. Lines are optical fibers, SW is an electro-optic switch, PC is a polarization controller and PBSC is a polarization beam splitter/combiner.}
\end{figure}
The gate is shown on~FIG.~\ref{fig:polarization-gate}. First, a single photon in a time-bin qubit state is incident from the left. The input polarization, say horizontal (H), is chosen to maximize the transmission through the fast interferometric electro-optic 1$\times$2 switch (SW) that routes the $\ket{s}$ and the $\ket{l}$ components to the lower and upper output branches res\-pec\-ti\-vely. The polarization of the lower branch is flipped to vertical (V) by a polarization controller (PC) and it is also delayed by $\Delta t$. This delay synchronizes the $\ket{s}$ and $\ket{l}$ components at the inputs of the polarizing beam splitter/combiner (PBSC). The latter reflects and transmits V and H polarizations, hence, the time-bin qubit is converted to a single-rail polarization qubit according to $\ket{s} \rightarrow \ket{V}$ and $\ket{l} \rightarrow \ket{H}$. Then, the output of the PBSC is fed into an all-fiber polarization controller that transforms any polarization to any other with negligible loss. After the rotation, the polarization qubit is converted back to an H~polarized time-bin qubit using the inverse of the operation done in the first section. Therefore, this gate can implement all unitary operations on a single time-bin qubit.

The switches are the only active components. The faster the switching speed, the smaller is the required separation between the $\ket{s}$ and $\ket{l}$ components, thus, the smaller is the required path length difference $cn\Delta t$, where $n$ is the effective refraction index of the fiber. With current technology, electro-optic material allows for switching speeds of at least 10~GHz, hence, the path length difference can be set at about 2~cm or less in standard single mode fiber. This is crucial since the relative phase between the branches has to be stabilized in temperature. In this case, a path length difference of 2~cm with 1~m long arms would require the temperature to be stable within 1~tenth of a degree Kelvin at room temperature, which is not difficult to achieve in the laboratory. 
The two PBSC can be fabricated with fused fiber couplers and therefore are practically lossless \cite{Diet05}. The major limiting factor is the insertion loss of the switches caused by the mode mismatch between the fiber core and the waveguide of the switch. This loss can be lowered down to about 1.5~dB for a total of 3~dB. 

\begin{figure}[!h]
\includegraphics[scale=0.7]{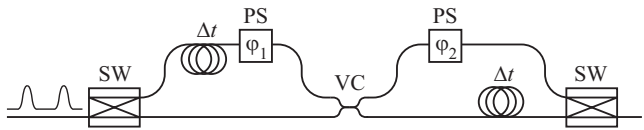}
\caption{\label{fig:dual-rail-gate} Experimental layout of the time-bin gate using dual-rail representation. Lines are optical fibers, SW is an electro-optic switch, PS is a phase shifter and VC is a variable coupler (beam splitter).}
\end{figure}

The second scheme we propose works by converting the time-bin qubit to a dual-rail qubit for processing~\cite{CY95}. The gate is pictured in~FIG.~\ref{fig:dual-rail-gate} and works as follows. A time-bin qubit is incident on a 1$\times$2 optical switch (SW) that routes the $\ket{s}$ and the $\ket{l}$ components to the upper and lower branches res\-pec\-ti\-vely. 
Next, the upper branch is delayed by $\Delta t$ to synchronize $\ket{s}$ with $\ket{l}$ and consequently, the photon is converted to the dual-rail qubit basis: $\ket{s} \rightarrow \ket{1}_u\ket{0}_l$ and $\ket{l} \rightarrow \ket{0}_u\ket{1}_l$, where the labels $u$ and $l$ stand for the upper and lower branches respectively. The two components of the single photon interfere classically at the variable coupler (VC) set to a specific coupling ratio to implement a given gate. This, supplemented with two phase shifters (PS) on the upper branch, can implement any unitary operation on the dual-rail qubit. The remaining part of the gate converts the dual-rail qubit back to a single-rail time-bin qubit.

The switches are the only active components so the speed and temperature requirements are the same as the other scheme. Moreover, stable and precise phase shifters can be made lossless using piezoactuators glued on the fiber. A lossless variable coupler can also be implemented by applying stress on an all-fiber 50/50 coupler to select the coupling ratio \cite{PBSC85}. Therefore, the only important sources of loss are the switches and again, it can be lowered down to about 3~dB in total.

We have proposed two experimental schemes to implement arbitrary unitary operations on single time-bin qubits using fiber optics components available with current technology. Future directions include experimental realization of the propositions and generalization to operations on time-bin qutrits and qudits. 

\bibliography{LEOS}

\begin{thebibliography}{13}
\expandafter\ifx\csname natexlab\endcsname\relax\def\natexlab#1{#1}\fi
\expandafter\ifx\csname bibnamefont\endcsname\relax
  \def\bibnamefont#1{#1}\fi
\expandafter\ifx\csname bibfnamefont\endcsname\relax
  \def\bibfnamefont#1{#1}\fi
\expandafter\ifx\csname citenamefont\endcsname\relax
  \def\citenamefont#1{#1}\fi
\expandafter\ifx\csname url\endcsname\relax
  \def\url#1{\texttt{#1}}\fi
\expandafter\ifx\csname urlprefix\endcsname\relax\def\urlprefix{URL }\fi
\providecommand{\bibinfo}[2]{#2}
\providecommand{\eprint}[2][]{\url{#2}}

\bibitem[{\citenamefont{Bennett and Brassard}(1984)}]{BB84}
\bibinfo{author}{\bibfnamefont{C.~H.} \bibnamefont{Bennett}} \bibnamefont{and}
  \bibinfo{author}{\bibfnamefont{G.}~\bibnamefont{Brassard}},
  \bibinfo{journal}{Proc. of the IEEE Int. conf. on Computers, Systems \&
  Signal Processing, Bangalore, India (IEEE, New York)} pp.
  \bibinfo{pages}{175--179} (\bibinfo{year}{1984}).

\bibitem[{\citenamefont{Briegel et~al.}(1998)\citenamefont{Briegel, Dur, Cirac,
  and Zoller}}]{BDCZ98}
\bibinfo{author}{\bibfnamefont{H.-J.} \bibnamefont{Briegel}},
  \bibinfo{author}{\bibfnamefont{W.}~\bibnamefont{Dur}},
  \bibinfo{author}{\bibfnamefont{J.~I.} \bibnamefont{Cirac}}, \bibnamefont{and}
  \bibinfo{author}{\bibfnamefont{P.}~\bibnamefont{Zoller}},
  \bibinfo{journal}{Phys. Rev. Lett.} \textbf{\bibinfo{volume}{81}},
  \bibinfo{pages}{5932} (\bibinfo{year}{1998}).

\bibitem[{\citenamefont{Bell}(1964)}]{Bell64}
\bibinfo{author}{\bibfnamefont{J.~S.} \bibnamefont{Bell}},
  \bibinfo{journal}{Rev. Mod. Phys.} \textbf{\bibinfo{volume}{38}},
  \bibinfo{pages}{447} (\bibinfo{year}{1964}).

\bibitem[{\citenamefont{Bennett et~al.}(1993)\citenamefont{Bennett, Brassard,
  Cr\'epeau, Jozsa, Peres, and Wootters}}]{BBCJPW93}
\bibinfo{author}{\bibfnamefont{C.~H.} \bibnamefont{Bennett}},
  \bibinfo{author}{\bibfnamefont{G.}~\bibnamefont{Brassard}},
  \bibinfo{author}{\bibfnamefont{C.}~\bibnamefont{Cr\'epeau}},
  \bibinfo{author}{\bibfnamefont{R.}~\bibnamefont{Jozsa}},
  \bibinfo{author}{\bibfnamefont{A.}~\bibnamefont{Peres}}, \bibnamefont{and}
  \bibinfo{author}{\bibfnamefont{W.~K.} \bibnamefont{Wootters}},
  \bibinfo{journal}{Phys. Rev. Lett.} \textbf{\bibinfo{volume}{70}},
  \bibinfo{pages}{1895} (\bibinfo{year}{1993}).

\bibitem[{\citenamefont{Brendel et~al.}(1999)\citenamefont{Brendel, Gisin,
  Tittel, and Zbinden}}]{BGTZ99}
\bibinfo{author}{\bibfnamefont{J.}~\bibnamefont{Brendel}},
  \bibinfo{author}{\bibfnamefont{N.}~\bibnamefont{Gisin}},
  \bibinfo{author}{\bibfnamefont{W.}~\bibnamefont{Tittel}}, \bibnamefont{and}
  \bibinfo{author}{\bibfnamefont{H.}~\bibnamefont{Zbinden}},
  \bibinfo{journal}{Phys. Rev. Lett.} \textbf{\bibinfo{volume}{82}},
  \bibinfo{pages}{2594} (\bibinfo{year}{1999}).

\bibitem[{\citenamefont{Gisin et~al.}(2002)\citenamefont{Gisin, Ribordy,
  Tittel, and Zbinden}}]{GRTZ02}
\bibinfo{author}{\bibfnamefont{N.}~\bibnamefont{Gisin}},
  \bibinfo{author}{\bibfnamefont{G.}~\bibnamefont{Ribordy}},
  \bibinfo{author}{\bibfnamefont{W.}~\bibnamefont{Tittel}}, \bibnamefont{and}
  \bibinfo{author}{\bibfnamefont{H.}~\bibnamefont{Zbinden}},
  \bibinfo{journal}{Rev. of Mod. Phys.} \textbf{\bibinfo{volume}{74}},
  \bibinfo{pages}{145} (\bibinfo{year}{2002}).

\bibitem[{\citenamefont{Tittel and Weihs}(2001)}]{TW01}
\bibinfo{author}{\bibfnamefont{W.}~\bibnamefont{Tittel}} \bibnamefont{and}
  \bibinfo{author}{\bibfnamefont{G.}~\bibnamefont{Weihs}},
  \bibinfo{journal}{Quantum Inf. and Comp.} \textbf{\bibinfo{volume}{1}},
  \bibinfo{pages}{3} (\bibinfo{year}{2001}).

\bibitem[{\citenamefont{Marcikic et~al.}(2003)\citenamefont{Marcikic,
  de~Riedmatten, Tittel, Zbinden, and Gisin}}]{MRTZG03}
\bibinfo{author}{\bibfnamefont{I.}~\bibnamefont{Marcikic}},
  \bibinfo{author}{\bibfnamefont{H.}~\bibnamefont{de~Riedmatten}},
  \bibinfo{author}{\bibfnamefont{W.}~\bibnamefont{Tittel}},
  \bibinfo{author}{\bibfnamefont{H.}~\bibnamefont{Zbinden}}, \bibnamefont{and}
  \bibinfo{author}{\bibfnamefont{N.}~\bibnamefont{Gisin}},
  \bibinfo{journal}{Nature} \textbf{\bibinfo{volume}{421}},
  \bibinfo{pages}{509} (\bibinfo{year}{2003}).

\bibitem[{\citenamefont{Marcikic et~al.}(2004)\citenamefont{Marcikic,
  de~Riedmatten, Tittel, Zbinden, Legr\'e, and Gisin}}]{MRTZLG04}
\bibinfo{author}{\bibfnamefont{I.}~\bibnamefont{Marcikic}},
  \bibinfo{author}{\bibfnamefont{H.}~\bibnamefont{de~Riedmatten}},
  \bibinfo{author}{\bibfnamefont{W.}~\bibnamefont{Tittel}},
  \bibinfo{author}{\bibfnamefont{H.}~\bibnamefont{Zbinden}},
  \bibinfo{author}{\bibfnamefont{M.}~\bibnamefont{Legr\'e}}, \bibnamefont{and}
  \bibinfo{author}{\bibfnamefont{N.}~\bibnamefont{Gisin}},
  \bibinfo{journal}{Phys. Rev. Lett.} \textbf{\bibinfo{volume}{93}},
  \bibinfo{pages}{180502} (\bibinfo{year}{2004}).

\bibitem[{\citenamefont{Fasel et~al.}(2004)\citenamefont{Fasel, Gisin, Ribordy,
  and Zbinden}}]{FGRZ04}
\bibinfo{author}{\bibfnamefont{S.}~\bibnamefont{Fasel}},
  \bibinfo{author}{\bibfnamefont{N.}~\bibnamefont{Gisin}},
  \bibinfo{author}{\bibfnamefont{G.}~\bibnamefont{Ribordy}}, \bibnamefont{and}
  \bibinfo{author}{\bibfnamefont{H.}~\bibnamefont{Zbinden}},
  \bibinfo{journal}{Eur. Phys. J. D} \textbf{\bibinfo{volume}{30}},
  \bibinfo{pages}{143} (\bibinfo{year}{2004}).

\bibitem[{\citenamefont{Aub\'e et~al.}(2005)\citenamefont{Aub\'e, Boucon,
  Burgoyne, Daxhelet, Dupuis, Godbout, Gonthier, Lacroix, Leclerc, Seguin
  et~al.}}]{Diet05}
\bibinfo{author}{\bibfnamefont{M.}~\bibnamefont{Aub\'e}},
  \bibinfo{author}{\bibfnamefont{A.}~\bibnamefont{Boucon}},
  \bibinfo{author}{\bibfnamefont{B.}~\bibnamefont{Burgoyne}},
  \bibinfo{author}{\bibfnamefont{X.}~\bibnamefont{Daxhelet}},
  \bibinfo{author}{\bibfnamefont{A.}~\bibnamefont{Dupuis}},
  \bibinfo{author}{\bibfnamefont{N.}~\bibnamefont{Godbout}},
  \bibinfo{author}{\bibfnamefont{F.}~\bibnamefont{Gonthier}},
  \bibinfo{author}{\bibfnamefont{S.}~\bibnamefont{Lacroix}},
  \bibinfo{author}{\bibfnamefont{M.-E.} \bibnamefont{Leclerc}},
  \bibinfo{author}{\bibfnamefont{F.}~\bibnamefont{Seguin}},
  \bibnamefont{et~al.}, \bibinfo{journal}{Photons: The Technical Review of the
  Canadian Institute for Photonics Innovations, Spring 2005}
  (\bibinfo{year}{2005}).

\bibitem[{\citenamefont{Chuang and Yamamoto}(1995)}]{CY95}
\bibinfo{author}{\bibfnamefont{I.}~\bibnamefont{Chuang}} \bibnamefont{and}
  \bibinfo{author}{\bibfnamefont{Y.}~\bibnamefont{Yamamoto}},
  \bibinfo{journal}{Phys. Rev. A} \textbf{\bibinfo{volume}{52}},
  \bibinfo{pages}{3489} (\bibinfo{year}{1995}).

\bibitem[{\citenamefont{Kawasaki et~al.}(1985)\citenamefont{Kawasaki, Kawachi,
  Hill, and Johnson}}]{PBSC85}
\bibinfo{author}{\bibfnamefont{B.~S.} \bibnamefont{Kawasaki}},
  \bibinfo{author}{\bibfnamefont{M.}~\bibnamefont{Kawachi}},
  \bibinfo{author}{\bibfnamefont{K.~O.} \bibnamefont{Hill}}, \bibnamefont{and}
  \bibinfo{author}{\bibfnamefont{D.~C.} \bibnamefont{Johnson}},
  \emph{\bibinfo{title}{Optical fiber coupler with tunable coupling ratio and
  method of making}}, \bibinfo{howpublished}{US Patent 4763977}
  (\bibinfo{year}{1985}).

\end{thebibliography}

\end{document}